\documentclass[12pt]{article}
\usepackage{graphicx}
\usepackage{amssymb}
\usepackage{amsmath}
\usepackage{bm}
\usepackage{cite}

\newcommand{\bear}{\begin{array}}  
\newcommand {\eear}{\end{array}}
\newcommand{\bea}{\begin{eqnarray}}   
\newcommand{\eea}{\end{eqnarray}}
\newcommand{\beq}{\begin{eqnarray}}   
\newcommand{\eeq}{\end{eqnarray}}
\newcommand{\bef}{\begin{figure}}  \newcommand 
{\eef}{\end{figure}}
\newcommand{\bec}{\begin{center}}  \newcommand 
{\eec}{\end{center}}

\newcommand{\1}{\mbox{1}\hspace{-0.25em}\mbox{l}}

\begin{document}

\begin{titlepage}

\begin{flushright}
IPMU13-0026  \\
\end{flushright}

\vskip 1.35cm
\begin{center}

{\large 
{\bf 
Two-loop Renormalization Factors \\of Dimension-six 
Proton Decay Operators \\
in the Supersymmetric Standard Models}}
\vskip 1.2cm

Junji Hisano$^{a,b}$, 
Daiki Kobayashi$^a$,
Yu Muramatsu$^a$,
and
Natsumi Nagata$^{a,c}$\\

\vskip 0.4cm

{\it $^a$Department of Physics,
Nagoya University, Nagoya 464-8602, Japan}\\
{\it $^b$Kavli IPMU, University of Tokyo, Kashiwa 277-8568, Japan}\\
{\it $^c$Department of Physics, 
University of Tokyo, Tokyo 113-0033, Japan}
\date{\today}

\vskip 1.5cm

\begin{abstract} 
 The renormalization factors of the dimension-six effective operators
 for proton decay are evaluated at two-loop level in the supersymmetric
 grand unified theories. For this purpose, we use the previous results
 in which the quantum corrections to the effective K\"{a}hler potential
 are evaluated at two-loop level. Numerical values for the factors are
 presented in the case of the minimal supersymmetric SU(5) grand unified
 model. We also derive a simple formula for the one-loop renormalization
 factors for any higher-dimensional operators in the K\"{a}hler
 potential, assuming that they are induced by the gauge interactions.

\end{abstract}

\end{center}
\end{titlepage}

\section{Introduction}

Discovery of the Higgs boson \cite{:2012gk, :2012gu} may suggest
the existence of supersymmetry (SUSY). The supersymmetric theories may
accommodate the hierarchical structure with the great desert
naturally. Searches for rare processes, such as proton decay, would be
one of methods to access the physics beyond the supersymmetric
standard models (SUSY SMs). The processes are dictated with the effective
higher-dimensional operators. When comparing the prediction with the
observation precisely, we need to include the radiative corrections
correctly. 

The realization of the gauge coupling unification strongly motivates us
to study the supersymmetric grand unified theories (SUSY GUTs)
\cite{Witten:1981nf,Dimopoulos:1981yj, Dimopoulos:1981zb,
Sakai:1981gr}. 
In the theories, proton decay is induced by the exchanges of the
colored Higgs multiplets and the $X$ gauge bosons, which yield the
baryon and lepton number non-conserving interactions. They are expressed
in terms of the dimension-five and -six effective operators,
respectively. It is found that the former interactions in general give
rise to dominant channels for proton decay, such as $p\to K^+\bar{\nu}$
\cite{Sakai:1981pk, Weinberg:1981wj}. However, the current experimental
limits on the channel, $\tau(p\to K^+ \bar{\nu})>3.3\times 10^{33}$~yrs
\cite{Miura:2010zz}, are so severe that the contribution of the
dimension-five operators is required to be suppressed by a certain
mechanism; otherwise the model is excluded just as the case of the
minimal SUSY SU(5) GUT unless the SUSY particles in the SUSY SM are much heavier than the weak scale\cite{Goto:1998qg, Murayama:2001ur}. A variety of
such mechanisms have been proposed. For example, the Peccei-Quinn
symmetry \cite{Peccei:1977hh} would be exploited for the purpose. The
$R$ symmetry also plays a role in suppressing the dimension-five proton
decay in the models with extra dimensions\cite{Hall:2001pg}.
With such a suppression mechanism imposed, the dimension-six operators
in turn become dominant. In this case, the main decay mode is the $p\to
\pi^0 e^+$ channel; the present experimental limit on its lifetime is
given by $\tau(p\to \pi^0 e^+)> 1.29\times 10^{34}$~yrs \cite{:2012rv}.
Since in the SUSY GUTs the GUT scale $M_{\rm GUT}$ is
relatively high, {\it i.e.}, $M_{\rm GUT}\sim 2\times 10^{16}~{\rm GeV}$, the
predicted proton lifetime usually evades the experimental limit.
However, the consequence might be altered if there exist extra particles
in the intermediate scale. 
With such particles belonging to a representation of the grand unified
group, the gauge coupling unification is still achieved, while its value
at the unified scale turns out to be enhanced. Then, the proton lifetime
is considerably reduced due to the large gauge coupling
\cite{Hisano:2012wq}.

In order to study such possibilities based on the
proton decay experiments, it is important to make a precise prediction
for the decay rate. To that end, we need to determine the effects of
the dimension-six operators, which are generated at the GUT scale, on the
low-energy physics by using the renormalization group equations (RGEs).
Indeed, there have been several literature in which the renormalization
factors for the effective operators are evaluated. In
Ref.~\cite{Nihei:1994tx}, the long-distance QCD corrections are computed
at two-loop level. For the short-distance factors, on the other hand, only
the one-loop calculation is carried out in Ref.~\cite{Munoz:1986kq} in
the SUSY SM. 

In this Letter, therefore, we evaluate the renormalization factors of
the dimension-six operators at two-loop level in the presence of the
supersymmetry. In the calculation, we use the results for the
two-loop corrections to the effective K\"{a}hler potential given in
Ref.~\cite{Nibbelink:2005wc} Since in the SUSY GUTs, the most of the
intermediate energy scales are supersymmetric, the short-distance
renormalization factors are well approximated by those evaluated in
purely SUSY theory. Thus, combined with the long-distance effects
given in Ref.~\cite{Nihei:1994tx}, our results offer a tool for making
a prediction of the proton decay rate with accuracy of two-loop level.

We also derive a simple formula for the one-loop level renormalization
factors of any higher-dimensional operators in the K\"{a}hler potential,
when including only the gauge interaction contributions. It is
applicable to other observables, such as the neutron-antineutron
oscillation \cite{HKMN}.

This Letter is organized as follows: in Sec.~\ref{dim6}, we first write
down the dimension-six effective operators in terms of the superfield
notation. Our notations and conventions are also
shown in the section. Then, in the subsequent section, we
describe a way of calculating the renormalization factors of the
operators by using the effective K\"{a}hler potential, and present the
results for the computation. In Sec.~\ref{results}, the comparison of
the one- and two-loop renormalization factors is discussed in the
minimal SUSY SU(5) GUT. Section \ref{conclusion} is devoted to
conclusion and discussion.

\section{Dimension-six effective operators}
\label{dim6}

To begin with, we write the dimension-six effective operators for proton
decay in a SUSY and gauge invariant manner with superspace notation:  
\begin{align}
{\cal O}^{(1)}&=\int d^2\theta d^2\bar{\theta}
\epsilon_{abc}\epsilon_{ij}
(\overline{U}^\dagger)^a(\overline{D}^\dagger)^b 
e^{-\frac{2}{3}g_Y V_1}
(e^{2g_3V_3}Q_i)^c L_j~,
\nonumber \\
{\cal O}^{(2)} &=\int d^2\theta d^2\bar{\theta}
\epsilon_{abc}\epsilon_{ij}\overline{E}^\dagger e^{\frac{2}{3}g_Y
 V_1}(e^{-2g_3V_3}\overline{U}^\dagger)^aQ^b_iQ^c_j~,
\label{effective_operators}
\end{align}
where all the chiral superfields correspond to left-handed fermions, and
$V_1$ and $V_3$ are the U(1)$_Y$ and SU(3)$_C$ vector superfields
with the gauge coupling constants $g_Y$ and $g_3$,
respectively. The subscripts $i,j$, are the SU(2)$_L$ indices, while
$a,b,c$ are the color indices. Furthermore, we omit the generation
indices for simplicity.

The relationship between bare and renormalized operators is written in
the following form:
\begin{equation}
 {\cal O}^{(I)}_B=Z^{(I)}{\cal O}^{(I)},~~~~~~(I=1,2)~,
\end{equation}
where the subscript $B$ indicates the operator is bare.
Then, the Wilson coefficients $C^{(I)}$ for the operators ${\cal
O}^{(I)}$ obey the differential equations,
\begin{equation}
 \mu \frac{d}{d\mu}C^{(I)}(\mu)=\gamma_{{\cal {O}}^{(I)}}C^{(I)}(\mu)~,
\end{equation}
with $\gamma_{{\cal O}^{(I)}}$ the anomalous dimensions for the operators
defined as
\begin{equation}
 \gamma_{{\cal O}^{(I)}}\equiv \mu \frac{d}{d\mu}\ln Z^{(I)}~.
\end{equation}

The anomalous dimensions are obtained by analyzing the vertex functions
(or the effective action) in which the operators are inserted. 
Since we now deal with the dimension-six operators which contain four
chiral or anti-chiral superfields, it is sufficient to consider the
four-point vertex functions which include the corresponding external
superfields. Their renormalization group equations (RGEs) are given as
\begin{equation}
 \biggl[
\mu \frac{\partial}{\partial \mu}+\beta_\alpha\frac{\partial}
{\partial g_\alpha}
-\sum_{i}\gamma_i +\gamma_{{\cal O}^{(I)}}
\biggr]\Gamma_{{\cal O}^{(I)}}=0~.
\label{RGE_gen}
\end{equation}
Here, $\Gamma_{{\cal O}^{(I)}}$ are the four-point vertex
functions with an insertion of the operators ${\cal O}^{(I)}$. The gauge
coupling constants and their beta functions are denoted by $g_\alpha$ and
$\beta_\alpha$, respectively, and the sum over each gauge group is
implicit. Further, $\gamma_i$ shows the anomalous dimension of each
superfield contained in the operators. From now on, we often omit the
superscript $(I)$ for brevity.

\section{Renormalization factors}
\label{renormalization}

In this section, we present the formulae for the renormalization
factors. They are derived from the effective K\"{a}hler potential given
in Ref.~\cite{Nibbelink:2005wc}. In the calculation, the dimensional
reduction scheme ($\overline{\rm DR}$) \cite{Siegel:1979wq} is employed
for the regularization.
We first obtain the one-loop results and confirm the results in
Ref.~\cite{Munoz:1986kq} in the former subsection. 
Then, in the latter subsection, we evaluate the two-loop contribution.

\subsection{One-loop}
\label{1loop}

Let us first evaluate the vertex functions at one-loop level. For this
purpose, we use the results in Ref.~\cite{Nibbelink:2005wc}, where the
effective K\"{a}hler potential for generic four-dimensional $N=1$ SUSY
theories is computed up to two-loop level. According to the results, the
one-loop correction\footnote{
This one-loop result is first derived in Ref.~\cite{Brignole:2000kg}.}
to the K\"{a}hler potential is given as 
\begin{equation}
 \Delta K_1=-\sum_{\alpha}\frac{1}{16\pi^2}{\tt Tr} M^{
2}_{C(\alpha)}\bigl(
2-\ln \frac{M_{C(\alpha)}^{2}}{\bar{\mu}^2}
\bigr)~,
\label{one_loop_K}
\end{equation}
where $\bar{\mu}^2\equiv 4\pi e^{-\gamma}\mu^2$ defines the
$\overline{\rm MS}$ renormalization scale, and the mass matrix
$M_{C(\alpha)}^{2}$
is defined by
\begin{equation}
 (M^{2}_{C(\alpha)})_{AB}\equiv 2g_\alpha^2\bar{\phi}_a(T^{(\alpha)}_A)^a_{~b}
G^b_{~c}(T_B^{(\alpha)})^c_{~d}\phi^d~,
\end{equation}
with $\phi$ the background for the chiral superfield $\Phi$ and $G^a_{~b}$
the K\"{a}hler metric 
\begin{equation}
 G^a_{~b}\equiv \frac{\partial^2}{\partial\bar{\phi}_a\partial\phi^b}
K(\bar{\phi}, \phi)~.
\label{Kahler_metric}
\end{equation}
In Eq.~\eqref{one_loop_K}, ${\tt Tr}$ denotes the trace over the adjoint
representation of a gauge group whose coupling constant is
$g_\alpha$ and generators are given by $T^{(\alpha)}_A
$. Moreover, in the following calculation, we only take the gauge
interactions into account, {\it i.e.}, we neglect the
superpotential.\footnote{
Experimental constraints on the effective operators in
Eq.~\eqref{effective_operators} are particularly severe when the
external lines of the operators are of the first and/or second
generations. In such a case, the size of the Yukawa couplings are
negligible. 
} 

In order to obtain the renormalization factors for the higher-dimensional
effective operators, we consider the K\"{a}hler potential 
\begin{equation}
 K=\bar{\phi}_a\phi^a+C{\cal O}+C{\cal O}^\dagger~,
\label{kahler}
\end{equation}
with $C$ the Wilson coefficient of the operator ${\cal O}$.
In this case, the K\"{a}hler metric reads
\begin{equation}
 G^a_{~b}=\delta^a_b+C{\cal O}^a_{~b}+C{\cal O}^{\dagger a}_{~~b}~,
\end{equation}
with ${\cal O}^a_{~b}\equiv \partial^2{\cal O}/\partial\bar{\phi}_a
\partial \phi^b$.
By substituting the above equations to Eq.~\eqref{one_loop_K}, we have
\begin{equation}
\Delta K_1=-\sum_{\alpha}\frac{g^2_\alpha}{16\pi^2}2
(1+\ln\bar{\mu}^2)
[C_\alpha (a)\bar{\phi}_a\phi^a+\{
C(\bar{\phi}T^{(\alpha)}_A)_a{\cal O}^a_{~b}
(T_A^{(\alpha)}\phi)^b
+{\rm h.c.}\}
]~,
\label{K_1loop}
\end{equation} 
where $C_\alpha(i)$ are the quadratic Casimir group theory invariants
for the superfield $\Phi_i$, defined in terms of the Lie algebra
generators $T_A$ by
$(T_A^{(\alpha)}T_A^{(\alpha)})^a_b=C_\alpha(i)\delta^a_b$. Further, we
keep only the terms up to the first order with respect to the Wilson
coefficient, $C$, and do not show the terms including the logarithmic
dependence on the background fields, which are not relevant to the
present calculation.
At the first order in the perturbation theory, the RGE \eqref{RGE_gen}
then leads to 
\begin{equation}
 \gamma_{\cal O}^{(1)}{\cal O}=\sum_{i}\gamma^{(1)}_i{\cal O}
+\sum_{\alpha}\frac{g^2_\alpha}{16\pi^2}4(\bar{\phi}T^{(\alpha)}_A)_a{\cal
O}^a_{~b}(T_A^{(\alpha)}\phi)^b~.
\label{RGE_1loop}
\end{equation}
Here, the superscript $(1)$ of the anomalous dimensions denotes that they
are evaluated at one-loop level. In supersymmetric theories,
$\gamma^{(1)}_i$ is given as\footnote{
The anomalous dimension of fields, $\gamma_i$, may be also derived
at one- and two-loop levels from the effective K\"{a}hler potential
derived in Ref.~\cite{Nibbelink:2005wc} in the similar way. See the
first term in Eq.~\eqref{K_1loop}. 
}
\begin{equation}
 \gamma^{(1)}_i=-2\sum_{\alpha}C_\alpha(i) \frac{g_\alpha^2}{16\pi^2}~.
\label{1_anom}
\end{equation}

Now we evaluate the second term in Eq.~\eqref{RGE_1loop}. To that end,
we analyze the structure of the term on a general basis in order to
derive the formula for the one-loop renormalization factor of any
operator. Consider the following operator which contains an arbitrary
number of both chiral and anti-chiral superfields and is singlet under a
given global symmetry $G$ as a whole: 
\begin{equation}
{\cal O}=\bar{\lambda}_a^{i_1\dots i_m}\lambda ^a_{j_1\dots j_n}
 \overline{\Phi}_{i_1}\dots  \overline{\Phi}_{i_m}\Phi^{j_1}\dots
\Phi^{j_n}~.
\end{equation}
Here, the coefficients $\lambda ^a_{j_1\dots j_n}$ and
$\bar{\lambda}_a^{i_1\dots i_m}$ make the set of superfields $G$
singlet. When $G$ is localized (gauged), the operator invariant under
both supersymmetry and the gauge symmetry is 
\begin{equation}
 \int d^2\theta d^2\bar{\theta}~(\bar{\lambda}_a^{i_1\dots i_m}
 \overline{\Phi}_{i_1}\dots  \overline{\Phi}_{i_m})
\bigl[e^{2gV^A_G T^A}\bigr]^a_{~b}
(\lambda ^b_{j_1\dots j_n}\Phi^{j_1}\dots\Phi^{j_n})~,
\end{equation}
where $g$ and $V_G^A$ are the coupling constant and the gauge
vector superfields of the gauge group $G$, respectively. Moreover, $T^A$
are assumed to be the generators for an irreducible representation, which are relevant to the transformation properties of the composite chiral
superfield $\Phi_{j_1}\dots\Phi_{j_n}$; under the gauge transformation,
$\Phi_{j_1}\dots\Phi_{j_n}$ is transformed as
\begin{equation}
(\lambda ^a_{j_1\dots j_n}\Phi^{j_1}\dots\Phi^{j_n})
\to (e^{ig\Lambda^A T^A}) ^a_{~b}
(\lambda ^b_{j_1\dots j_n}\Phi^{j_1}\dots\Phi^{j_n})~,
\end{equation}
with $\Lambda^A$ any chiral superfields.  Further,
we write the generators for each chiral superfield $\Phi$ as
$t^A$, {\it i.e.}, $\Phi^j\to (e^{ig\Lambda^A
t^A})^j_{~j^\prime}\Phi^{j^\prime}$. 
Then, since the coefficients $\lambda ^a_{j_1\dots j_n}$ and
$\bar{\lambda}_a^{i_1\dots i_m}$ assemble the transformation properties
of each chiral superfield into that of the composite operator $\lambda
^a_{j_1\dots j_n}\Phi^{j_1}\dots\Phi^{j_n}$, it follows that
\begin{equation}
 (T^A)^a_{~b}\lambda ^b_{j_1\dots j_n}
=\lambda ^a_{j_1^\prime j_2\dots j_n}(t^A)^{j_1^\prime}_{~j_1}+\dots +
\lambda ^a_{j_1\dots j_{n-1} j_n^\prime}(t^A)^{j_n^\prime}_{~j_n}~,
\end{equation}
and similarly for the anti-chiral superfields,
\begin{equation}
 \bar{\lambda}_b^{i_1\dots i_m}(T^A)^b_{~a}
=(t^A)^{i_1}_{~i_1^\prime} ~\bar{\lambda}_b^{i_1^\prime i_2\dots i_m}
+\dots +(t^A)^{i_m}_{~i_m^\prime}
~\bar{\lambda}_b^{i_1 \dots i_{m-1} i_m^\prime}~.
\end{equation}
These expressions imply that $\lambda ^a_{j_1\dots j_n}$ and
$\bar{\lambda}_a^{i_1\dots i_m}$ are invariant tensors under $G$.
By using the relations, we now evaluate the second term in
Eq.~\eqref{RGE_1loop}. It goes as follows:
\begin{align}
(\bar{\phi}t^A)_a{\cal O}^a_{~b}(t^A\phi)^b
&=[(t^A)^{i_1}_{~i_1^\prime} ~\bar{\lambda}_b^{i_1^\prime i_2\dots i_m}
+\dots +(t^A)^{i_m}_{~i_m^\prime}
~\bar{\lambda}_b^{i_1 \dots i_{m-1} i_m^\prime}]\nonumber\\
&\times
[\lambda ^a_{j_1^\prime j_2\dots j_n}(t^A)^{j_1^\prime}_{~j_1}+\dots +
\lambda ^a_{j_1\dots j_{n-1} j_n^\prime}(t^A)^{j_n^\prime}_{~j_n}]
\bar{\phi}_{i_1}\dots\bar{\phi}_{i_m}\phi^{j_1}\dots \phi^{j_n}
\nonumber \\
& =\bar{\lambda}_b^{i_1\dots i_m}(T^A)^b_{~a}
 (T^A)^a_{~c}\lambda ^c_{j_1\dots j_n}
\bar{\phi}_{i_1}\dots\bar{\phi}_{i_m}\phi^{j_1}\dots \phi^{j_n}\nonumber \\
&=C_G^{\rm comp}~{\cal O}~,
\end{align}
where $C_G^{\rm comp}$ is defined by $T^A T^A=C^{\rm comp}_G \1$; it
corresponds to the Casimir invariant for the composite chiral superfield
$\lambda ^a_{j_1\dots j_n}\Phi^{j_1}\dots\Phi^{j_n}$. Substituting the
expression into Eq.~\eqref{RGE_1loop}, we finally obtain a generic
formula for the one-loop renormalization factors of arbitrary 
operators: 
\begin{equation}
 \gamma^{(1)}_{\cal O}=\sum_{\alpha}\frac{g_\alpha^2}{16\pi^2}\biggl[
4C^{\rm comp}_\alpha - 2\sum_{i}C_\alpha (i)
\biggr]~,
\end{equation}
with $C^{\rm comp}_\alpha$ the Casimir invariants of the gauge group
$\alpha$ for the chiral part of the operators.

So far we have assumed that the set of chiral (anti-chiral)
superfields forms an irreducible representation. When it is reducible,
independent operators are formed. They are not mixed with each other at
one-loop level if only gauge interactions are effective. 

Now we apply the formula to the dimension-six effective operators for
proton decay  in Eq.~\eqref{effective_operators}. We find $C_3^{\rm
comp}=C_3(\square)=4/3$ in the case of SU(3)$_C$ and $C_2^{\rm comp}=0$
in the case of SU(2)$_L$ for both ${\cal O}^{(1)}$ and ${\cal O}^{(2)}$.
Here, $\square$ denotes the fundamental representation of the
corresponding group, and we have used $C_3(\square)=C_3(\overline{\square})$.
Note that the latter equation for SU(2)$_L$ follows from the fact that the
SU(2)$_L$ non-singlet superfields in the effective operators have the
same chirality and form an SU(2)$_L$ singlet. For U(1)$_Y$
contributions, on the other hand, we obtain different results for the
operators ${\cal O}^{(1)}$ and ${\cal O}^{(2)}$: $C_Y^{\rm comp} =
(Y_{Q}+Y_{L})^2$ for ${\cal O}^{(1)}$ and $C_Y^{\rm comp} =
(2Y_{Q})^2$ for ${\cal O}^{(2)}$.
As a result, by using these factors we obtain that 
\begin{equation}
 \gamma^{(1)}_{{\cal O}^{(I)}}=\sum_{\alpha=Y,2,3}\frac{g_\alpha^2}
{16\pi^2}\bigl[ \gamma^{(1)}_{{\cal O}^{(I)}}\bigr]_\alpha~,
\end{equation}
where
\begin{equation}
 \bigl[ \gamma^{(1)}_{{\cal O}^{(1)}}\bigr]_3=
\bigl[ \gamma^{(1)}_{{\cal O}^{(2)}}\bigr]_3=
-\frac{8}{3}~,
\end{equation}
\begin{equation}
 \bigl[ \gamma^{(1)}_{{\cal O}^{(1)}}\bigr]_2=
\bigl[ \gamma^{(1)}_{{\cal O}^{(2)}}\bigr]_2=
-3~,
\end{equation}
\begin{align}
 \bigl[ \gamma^{(1)}_{{\cal O}^{(1)}}\bigr]_Y&=-\frac{11}{9}~,\nonumber \\
 \bigl[ \gamma^{(1)}_{{\cal O}^{(2)}}\bigr]_Y&=-\frac{23}{9}~.
\end{align}
These results are totally consistent with those in
Ref.~\cite{Munoz:1986kq}.

\subsection{Two-loop}
\label{2loop}

Next, we discuss the two-loop level contribution. Again, we use the
results in Ref.~\cite{Nibbelink:2005wc}. The radiative corrections to
the K\"{a}hler potential at two-loop level are described by
\begin{align}
 \Delta K_2&=\frac{1}{2}R^{b~d}_{~a~c}{J}^{a~c}_{~b~d}(M^2)
-\sum_{\alpha}f^{(\alpha)}_{ABC}f^{(\alpha)}
_{DEF}{I}^{BDEAFC}(M_{V(\alpha)}^2) \nonumber \\
&-\sum_{\alpha}(GT^{(\alpha)}_A\phi)^b_{~;c}
(\bar{\phi}T^{(\alpha)}_B G)_a^{~;d}
H^{a~c~AB}_{~b~d}(M^2, M_{V(\alpha)}^2)~,
\label{two_loop_K}
\end{align}
with $f^{(\alpha)}_{ABC}$ the structure constants of the gauge group
$\alpha$. The mass functions and the geometric factors appear in
Eq.~\eqref{two_loop_K} are displayed in Appendix. By using them, we
readily obtain the two-loop corrections to the vertex functions. We
found from explicit calculation that the two-loop correction is not
given simply by the gauge transformation properties of the composite
chiral superfield in the operator and anomalous dimension of the
external fields, which is different from the one-loop ones. At present,
however, since the explicit derivations are quite complicated, we simply
give the final results and defer full details \cite{HKMN}.

The RGE in Eq.~\eqref{RGE_gen} at two-loop level is given as
\begin{equation}
 \mu\frac{\partial \Gamma_{\cal O}^{(2)}}{\partial\mu}
+\sum_{\alpha}\frac{1}{16\pi^2}b_\alpha
g_\alpha^3\frac{\partial}{\partial g_\alpha}\Gamma^{(1)}_{\cal O}
-\sum_{i}\gamma_i^{(1)}\Gamma^{(1)}_{\cal O}
-\sum_{i}\gamma_i^{(2)}\Gamma_{\cal O}^{(0)}
+\gamma_{\cal O}^{(1)}\Gamma_{\cal O}^{(1)}
+\gamma_{\cal O}^{(2)}\Gamma_{\cal O}^{(0)}=0~.
\label{RGE_2loop}
\end{equation}
Here, the subscripts (0--2) indicate the quantities are
evaluated at tree, one-loop, and two-loop level, respectively.
One-loop anomalous dimensions $\gamma_i^{(1)}$ are shown in
Eq.~\eqref{1_anom}, while the two-loop ones are given as
\cite{Martin:1993zk} 
\begin{equation}
 \gamma_i^{(2)}=\frac{1}{(16\pi^2)^2}
\sum_{\alpha,\beta}2g_\alpha^2C_\alpha(i)\bigl[g_\alpha^2b_\alpha 
\delta_{\alpha\beta}+2g_\beta^2C_\beta(i)\bigr]~.
\end{equation}
Here, $b_\alpha$ are the one-loop beta function coefficients for gauge
 coupling constants, given as $
 b_\alpha =\sum_{i}I_\alpha (i)-3C_\alpha (G)$ with
$C_\alpha (G)$ and $I_\alpha(i)$ the quadratic Casimir invariant for the
 adjoint representation of the group $\alpha$ and the
Dynkin index of the chiral multiplet $\Phi_i$, respectively.

From the RGE in Eq.~\eqref{RGE_2loop}, we now obtain the two-loop
anomalous dimensions for the effective operators. Again, we parametrize
them as follows:
\begin{align}
 \gamma^{(2)}_{{\cal O}^{(I)}}&=\frac{g_3^4}
{(16\pi^2)^2}\bigl[\gamma^{(2)}_{{\cal O}^{(I)}}\bigr]_{33} + \frac{g_2^4}
{(16\pi^2)^2}\bigl[\gamma^{(2)}_{{\cal O}^{(I)}}\bigr]_{22} + \frac{g_Y^4}
{(16\pi^2)^2}\bigl[\gamma^{(2)}_{{\cal O}^{(I)}}\bigr]_{YY} \nonumber\\
&+\frac{g_2^2 g_3^2}
{(16\pi^2)^2}\bigl[\gamma^{(2)}_{{\cal O}^{(I)}}\bigr]_{23} + \frac{g_Y^2 g_2^2}
{(16\pi^2)^2}\bigl[\gamma^{(2)}_{{\cal O}^{(I)}}\bigr]_{Y2} + \frac{g_Y^2 g_3^2}
{(16\pi^2)^2}\bigl[\gamma^{(2)}_{{\cal O}^{(I)}}\bigr]_{Y3}~.
\end{align}
Then, we have
\begin{equation}
\bigl[\gamma^{(2)}_{{\cal O}^{(1)}}\bigr]_{33}=
\bigl[\gamma^{(2)}_{{\cal O}^{(2)}}\bigr]_{33}=
 \frac{64}{3}+8b_3~,
\end{equation}
\begin{equation}
\bigl[\gamma^{(2)}_{{\cal O}^{(1)}}\bigr]_{22}=
\bigl[\gamma^{(2)}_{{\cal O}^{(2)}}\bigr]_{22}=
\frac{9}{2}+3b_2~,
\end{equation}
\begin{align}
\bigl[\gamma^{(2)}_{{\cal O}^{(1)}}\bigr]_{YY}= &
\frac{113}{54}+\frac{5}{3}b_Y~,\nonumber\\
\bigl[\gamma^{(2)}_{{\cal O}^{(2)}}\bigr]_{YY}= &
\frac{91}{18}+3b_Y~,
\end{align}
\begin{align}
 \bigl[\gamma^{(2)}_{{\cal O}^{(1)}}\bigr]_{23}&=12~,\nonumber \\
 \bigl[\gamma^{(2)}_{{\cal O}^{(2)}}\bigr]_{23}&=20~,
\end{align}
\begin{align}
 \bigl[\gamma^{(2)}_{{\cal O}^{(1)}}\bigr]_{Y2}&=2~,\nonumber \\
 \bigl[\gamma^{(2)}_{{\cal O}^{(2)}}\bigr]_{Y2}&=\frac{2}{3}~,
\end{align}
\begin{align}
 \bigl[\gamma^{(2)}_{{\cal O}^{(1)}}\bigr]_{Y3}&=\frac{68}{9}~,\nonumber \\
 \bigl[\gamma^{(2)}_{{\cal O}^{(2)}}\bigr]_{Y3}&=\frac{76}{9}~.
\end{align}

\section{Results}
\label{results}
In this section, we give the numerical results of the renormalization
factors in the minimal SUSY SU(5) GUT. The short-distance
renormalization factors $A_S^{(I)}$ are defined as the ratios of the
coefficients $C^{(I)}$ for the effective operators at the SUSY scale
$M_{\mathrm{SUSY}}$ to those at the GUT scale $M_{\mathrm{GUT}}$: 
\begin{equation}
A_S^{(I)} \equiv
 \frac{C^{(I)}(M_{\mathrm{SUSY}})}{C^{(I)}(M_{\mathrm{GUT}})}~, \qquad
 (I=1, 2)~, 
\end{equation}
where we assume $M_{\mathrm{SUSY}}=1$ $\mathrm{TeV}$ and
$M_{\mathrm{GUT}}=1.5\times 10^{16}$ $\mathrm{GeV}$.  
The numerical results  at one-loop level are given as
\begin{align}
\begin{split}
A_S^{(1)}(1\text{-loop}) &=  1.959~,\\
A_S^{(2)}(1\text{-loop}) &= 2.058~,
\end{split}
\end{align}
while  at two-loop level, we have found 
\begin{align}
\begin{split}
A_S^{(1)}(2\text{-loop}) &=  1.961~, \\
A_S^{(2)}(2\text{-loop}) &= 2.052~.
\end{split}
\end{align}
Here, we calculate the one-loop (two-loop) short-distance factors with
the one-loop (two-loop) renormalization equations for the gauge coupling
constants in the SUSY SM\cite{Martin:1993zk}. The numerical values of the unified gauge coupling constant at the one- and two-loop level are given as $\alpha_5(1\text{-loop})=0.03906$ and $\alpha_5(2\text{-loop})=0.03968$, respectively, where $\alpha_5$ is defined as $\alpha_5 \equiv  g_3^2(M_{\mathrm{GUT}})/4\pi$. The results are hardly affected by the uncertainty of the input parameters, {\it e.g.}, the SU(3) gauge coupling constant, $\alpha_s(m_Z)=0.1184(7)$\cite{Beringer:1900zz}. There is a cancellation among the two-loop corrections  since the signs of $\bigl[\gamma^{(2)}_{{\cal
O}^{(1)}}\bigr]_{33}$ and $\bigl[\gamma^{(2)}_{{\cal
O}^{(2)}}\bigr]_{33}$ are opposite to those of
        the other two-loop anomalous dimensions. Therefore, the numerical values at two-loop level hardly differ from the one-loop
        ones. Without cancellations, the significance of the two-loop contributions to the short-distance factors reaches a few percent of the one-loop ones.
\section{Conclusion and discussion}
\label{conclusion}
We have evaluated the short-distance renormalization factors for the
dimension-six proton decay operators at two-loop level with  the
effective K\"{a}hler potential. The procedure described in this Letter
is generic and applicable to any higher-dimensional operators. We get
the results $A_S^{(1)}(2\text{-loop}) =  1.961$ and $A_S^{(2)}(2
\text{-loop}) = 2.052$ in the minimal SUSY SU(5) GUT. We have found that
the two-loop contributions hardly change the renormalization factors evaluated
        at one-loop level.

Finally, we briefly comment on the extensions of the minimal SUSY GUT.
The gauge coupling constants at the GUT scale increase if there
exist extra particles in the intermediate scale. The two-loop effects
may be more significant in such cases. 
In addition, let us note that our results are only for the SU(3)$_C
\times$ SU(2)$_L\times$U(1)$_Y$ gauge interactions. If some new gauge
interactions exist below the GUT scale, we also need to evaluate the
contributions of the gauge interactions. Even for such theories, however,
it is possible to execute the prescription describe above to estimate
the renormalization factors by means of the effective K\"{a}hler
potential. 

In this Letter, we neglect the possible effects of the threshold
corrections from particles whose masses are around the GUT scale. 
Although the effects are model-dependent, to
complete the two-loop level calculation, we also need to evaluate such
corrections. We will discuss the issue on another occasion \cite{HKMN}.

\section*{Acknowledgments}

The work of NN is supported by Research Fellowships of the Japan
Society for the Promotion of Science for Young Scientists. The work of
JH is supported by Grant-in-Aid for Scientific research from the
Ministry of Education, Science, Sports, and Culture (MEXT), Japan,
No. 20244037, No. 20540252, No. 22244021 and No. 23104011, and also by
World Premier International Research Center Initiative (WPI
Initiative), MEXT, Japan.

\section*{Appendix}

Here, we show the explicit form of the mass functions as well as the
geometric factors given in Eq.~\eqref{two_loop_K}: 
\begin{equation}
 {J}^{a~c}_{~b~d}(M^2)=\frac{2}{(16\pi^2)^2}(\ln\bar{\mu}^2)
\sum_{\alpha,\beta}(M^2_\alpha G^{-1})^a_{~b} 
(M^2_\beta G^{-1})^c_{~d}~,
\end{equation}
\begin{align}
 &I^{ABCDEF}(M^2_{V(\alpha)})=-\frac{1}{2}\frac{g^2_\alpha}
{(16\pi^2)^2}(\ln\bar{\mu}^2)
\bigl[4(M_{V(\alpha)}^2)_{AB}\delta_{CD}\delta_{EF}
\nonumber \\
&
-\delta_{AB}(M_{V(\alpha)}^2\ln M_{V(\alpha)}^2)_{CD}
\delta_{EF}-\delta_{AB}\delta_{CD}
(M_{V(\alpha)}^2\ln M_{V(\alpha)}^2 )_{EF}
\bigr]+{\rm cycl.}~,
\end{align}
where the "cycl." denotes the cyclic permutations of the labels $AB, CD,
EF$, and
\begin{align}
& H^{a~c~AB}_{~b~d}(M^2, M_{V(\alpha)}^2)=
-\frac{g_\alpha^2}{(16\pi^2)^2}(\ln\bar{\mu}^2)\times\nonumber \\
& \biggl[
\sum_{\beta}\delta_{AB}\bigl\{
2(M^2_\beta G^{-1})^a_{~b}(G^{-1})^c_{~d}
+2(G^{-1})^a_{~b}(M_\beta^2G^{-1})^c_{~d} \nonumber \\
&-(G^{-1})^a_{~b}(M_\beta^2\ln \{M_\beta^2 \}G^{-1})^c_{~d}
-(M_\beta^2\ln \{M_\beta^2 \}G^{-1})^a_{~b}(G^{-1})^c_{~d}
\bigr\}
\nonumber \\
&+2(G^{-1})^a_{~b}(G^{-1})^c_{~d}(M_{V(\alpha)}^2)_{AB}
+(G^{-1})^a_{~b}(G^{-1})^c_{~d}(M_{V(\alpha)}^2
\ln  M_{V(\alpha)}^2)_{AB}
\biggr]~,
\end{align}
Here, we drop the terms independent of the scale $\mu$ or
containing two logarithms. The latter terms give rise to the logarithmic
terms after differentiation, which cancel other logarithmic terms in the
RGEs.  
The mass parameters are defined as
\begin{equation}
 (M^2_\alpha)^a_{~b}\equiv 2g^2_\alpha(T_A^{(\alpha)}\phi)^a(\bar{\phi}T_A
^{(\alpha)} G)_b~,
\end{equation}
and
\begin{equation}
 (M_{V(\alpha)}^2)_{AB}\equiv \frac{1}{2}\bigl[
(M^2_{C(\alpha)})_{AB}+(M^2_{C(\alpha)})_{BA}
\bigr]~.
\end{equation}
Further, $G^{-1}$ is inverse of the K\"{a}hler metric $G^a_{~b}$ defined
in Eq.~\eqref{Kahler_metric}, and the curvature $R^{a~c}_{~b~d}$ is
given by
\begin{equation}
 R^{a~c}_{~b~d}\equiv
\frac{\partial^2}{\partial\bar{\phi}_{a}\partial\phi^b}
G^{c}_{~d}-\biggl(\frac{\partial}{\partial \bar{\phi}_{a}}
G^{c}_e\biggr)(G^{-1})^e_{~f}
\biggl(\frac{\partial}{\partial \phi^b}G^{f}_{~d}\biggr)~.
\end{equation}
The third term in Eq.~\eqref{two_loop_K} includes the shorthand
notations, $(GT_A\phi)^b_{~;c}$ and $(\bar{\phi}T_BG)_a^{~;d}$, which are
defined as
\begin{align}
 (GT_A\phi)^a_{~;b}&\equiv
G^a_{~c}(T_A)^c_{~b}+\biggl(\frac{\partial}{\partial \phi^c}
G^a_{~b}\biggr)(T_A\phi)^c \nonumber\\
&=(T_A)^a_{~c}G^c_{~b}+(\bar{\phi}T_A)_c\biggl(
\frac{\partial}{\partial \bar{\phi}_c}G^a_{~b}
\biggr)\equiv (\bar{\phi}T_A G)_b^{~;a}~.
\end{align}
Here, the second line follows from the gauge invariance of the
K\"{a}hler potential.

{}

\end{document}